\documentclass[prd,twocolumn,amsmath,amssymb,showpacs,floatfix,superscriptaddress,nofootinbib,preprintnumbers]{revtex4}
\usepackage{graphicx}
\usepackage{epsfig}
\usepackage{bm}
\usepackage{amsfonts}

\newcommand{\la}{\lambda}
\newcommand{\ud}{\mathrm{d}}
\newcommand{\ve}{\varepsilon}

\begin{document}

\title{Non-minimal quintessence and phantom with nearly flat potentials}

\author{Gaveshna Gupta}
\affiliation{Center For Theoretical Physics, Jamia Millia Islamia,
New Delhi 110025, India}

\author{Emmanuel N. Saridakis }
\email{msaridak@phys.uoa.gr} \affiliation{Department of Physics,
University of Athens, GR-15771 Athens, Greece}

\author{Anjan A. Sen}
\email{anjan.ctp@jmi.ac.in} \affiliation{Center For Theoretical
Physics, Jamia Millia Islamia, New Delhi 110025, India}

\begin{abstract}
We investigate quintessence and phantom dark energy scenarios, in
which the scalar fields evolve in nearly flat potentials and are
non-minimally coupled to gravity. We show that all such models
converge to a common behavior and we provide the corresponding
approximate analytical expressions for $w(\Omega_\phi)$ and
$w(a)$. We find that non-minimal coupling leads to richer
cosmological behavior comparing to its minimal counterpart. In
addition, comparison with Baryon Acoustic Oscillation and latest
Supernovae data reveals that agreement can be established more
easily and with less strict constraints on the model parameters.
 \end{abstract}

 \maketitle

\section{Introduction}

According to recent cosmological observations, based on Supernovae
Ia  \cite{obs1} and cosmic microwave background radiation (CMBR)
\cite{cmbr} probes, as well as to WMAP data \cite{wmap}, the
universe is experiencing an accelerated expansion. In order to
explain this unexpected behavior, one can modify the gravitational
theory \cite{ordishov}, or construct various ``field'' models of
dark energy. The most studied models of the literature consider a
canonical scalar field (quintessence) \cite{quintess}, a phantom
field, that is a scalar field with a negative sign of the kinetic
term \cite{phant}, or the combination of quintessence and phantom
in a unified model named quintom \cite{quintom}.

In such field dark energy scenarios the potential choice plays a
central role in the determination of the cosmological evolution.
However, one can acquire potential-independent, general behavior
under the assumption of slow-roll conditions, which can be
embedded in a potential chosen to be nearly flat. These potentials
have been shown to present interesting cosmological features,
especially in the case where the dark energy equation-of-state
parameter $w$ is around $-1$. Although there is not a concrete
proof, there are many arguments indicating that in the sub-class
of ``field'' dark energy models, nearly flat potentials are
perhaps a natural and simple way of preserving $w\approx-1$,
either in quintessence \cite{quinteflat} or phantom \cite{Kujat}
case (although a sufficiently large Hubble friction could also
lead to such a $w$ for arbitrary potentials). Finally, we mention
that nearly flat potentials, which keep the variations of the
scalar fields from their initial to their present values small,
can also be efficient to avoid unknown quantum gravity effects
\cite{Huang}.

In \cite{Scherrer:2007pu} it was shown that in all quintessence
models with nearly flat potentials, the evolution of $w$ converges
to a common behavior and can be described by a unique expression,
which can be approximately given analytically in terms of its
present value together with the value of dark-energy density
parameter at present. In \cite{Scherrer:2008be} this result was
extended to phantom models, while the generalization to quintom
scenario was performed in \cite{Setare:2008sf} where a universal
expression for $w$ was also extracted, allowing for a crossing of
$-1$. Recently, a similar study with a minimally coupled,
tachyon-type scalar field was implemented in\cite{amna}, resulting
exactly at the same equation of state with the one obtained for a
standard minimally coupled scalar field
\cite{Scherrer:2007pu,Scherrer:2008be}.

On the other hand, dark energy models where the fields are
non-minimally coupled to gravity \cite{Uzan:1999ch,nonminimal}
have been shown to present significant cosmological features
\cite{liddle}. In our recent work  \cite{Sen:2009mc} we
investigated non-minimal quintessence with nearly-flat potential,
in the context of Brans-Dicke framework. Under a conformal
transformation it becomes a coupled quintessence model and a
universal expression for $w$ can be extracted, depending
additionally on the coupling parameter.

In the present work we are interested in studying both
quintessence and phantom scenarios with nearly-flat potential, in
the general non-minimal framework. The plan of the work is as
follows: In section \ref{model} we present the non-minimally
quintessence and phantom models and we extract the approximated
general solutions for the sub-class of  nearly flat potentials. In
section \ref{discuss} we compare our formulae with the exact
cosmological evolution and we provide the observational
constraints on the parameters of the model. Finally, section
\ref{concl} is devoted to the summary of the obtained results.

\section{Evolution of non-minimally  quintessence and
phantom with nearly flat potentials}\label{model}

Let us  construct quintessence and phantom models with the field
being non-minimally coupled to gravity. In order to incorporate
both scenarios in a general and unified way, in the following we
introduce the usual $\ve$-parameter, acquiring the value $+1$ for
quintessence, and $-1$ for the phantom case. Throughout the work
we consider a flat Robertson-Walker metric:
\begin{equation}\label{metric}
ds^{2}=dt^{2}-a^{2}(t)d\bf{x}^2,
\end{equation}
with $a$ the scale factor.

The action of a universe constituted of a non-minimally coupled
 field $\phi$ (canonical or phantom) is \cite{nonminimal}:{\small{
\begin{equation}
S=\int\ud^{4}x\sqrt{-g}\Bigg[\frac{1}{2\kappa^{2}}R
-\ve\frac{1}{2} \Big(\partial_{\mu}\phi\partial^{\mu}\phi+\xi
R\phi^{2}\Big) - V(\phi)+\mathcal{L}_m\Bigg] \label{action}
\end{equation}}}
where $\kappa^{2}=8\pi G$, $\xi$ is the non-minimal coupling
parameter, $R=6(\frac{\ddot{a}}{a}+\frac{\dot{a}^{2}}{a^{2}})$ is
the Ricci scalar and a dot denotes differentiation with respect to
cosmological time. The Friedmann equations read:
\begin{equation}
\frac{3}{\kappa^{2}}H^{2}=\rho_{\phi}+\rho_{m},
\end{equation}
\begin{equation}
\dot{H}=-\frac{\kappa^2}{2}\Big[(\rho_{\phi}+p_{\phi})+\rho_{m}(1+w_{m})\Big].
\end{equation}
In these expressions,
 $\rho_m$ is the matter energy density, with $w_m$ the matter
 equation-of-state parameter. $\rho_{\phi}$ and $p_{\phi}$ are
 respectively the energy density and pressure of the non-minimally coupled
scalar field, given by \cite{nonminimal,Uzan:1999ch}:
\begin{eqnarray}
\label{eq:6a}
 &&\rho_{\phi}= \ve\frac{1}{2}\dot{\phi}^{2} + V(\phi)
+
 \ve6 \xi H\phi\dot{\phi} + \ve3\xi H^{2}\phi^{2},\\
 &&p_{\phi}= \ve\frac{1}{2}(1-4\xi)\dot{\phi}^{2} - V(\phi) +  \ve2 \xi
H \phi\dot{\phi} -\nonumber\\
 \label{eq:6b}
 &&  -\ve2 \xi (1-6\xi)\dot{H}\phi^{2}
- \ve3\xi(1-8\xi)H^{2}\phi^{2} + 2\xi\phi V'(\phi).\ \ \
\end{eqnarray}
In such a scenario, dark energy is attributed to the scalar field,
and thus its equation-of-state  parameter reads:
\begin{equation}\label{EoS}
w_{\phi}=\frac{p_\phi}{\rho_\phi},
\end{equation}
where we can equivalently consider the barotropic variable $
\gamma_\phi\equiv w_\phi+1$.
 Finally, the equations close by considering the
evolution equation for the scalar field, which takes the form
\cite{nonminimal,Uzan:1999ch}:
\begin{equation}
\ddot{\phi}+3H\dot{\phi}+6\xi(\dot{H}+2H^2)\phi+\ve V'(\phi)=0.
\end{equation}

It proves convenient to introduce
\begin{eqnarray}
&&\rho_{\text{tot}}=\rho_{\phi}+\rho_{m}=\frac{3H^2}{\kappa^2}
\label{rhoeff}\\
&&p_{\text{tot}}=p_{\phi}+w_{m}\rho_{m}, \label{peff}
\end{eqnarray}
and thus to define
\begin{equation}
\label{eq:13}
w_{\text{tot}}=\frac{p_{\text{tot}}}{\rho_{\text{tot}}}  =
\frac{\kappa^{2}p_{\phi}}{3H^{2}}+w_{m}\Omega_{m},
\end{equation}
where $\Omega_{m}$ is the matter density parameter. Therefore, the
second Friedmann equation can be rewritten as
\begin{equation}
\dot{H}=-\frac{\kappa^2}{2}(\rho_{\text{tot}}+p_{\text{tot}}) =
-\frac{3}{2}H^{2}(1+w_{\text{tot}}). \label{Hdoteff}
\end{equation}
The variable  $w_{\text{tot}}$ is an intermediate one and will be
useful in order to eliminate $\dot{H}$ (using relation
(\ref{Hdoteff})) from all equations, since, contrary to the
minimal coupling case, now we face the presence of $\dot{H}$ in
the definition of the dark energy equation-of-state parameter
(\ref{EoS}).

At this stage, we desire to transform the aforementioned
cosmological system into an autonomous form
\cite{Copeland:1997et}. This will be achieved by introducing the
auxiliary variables:
\begin{eqnarray}
&&x=\frac{\kappa\dot{\phi}}{\sqrt{6}H} \nonumber\\
&&y=\frac{\kappa\sqrt{V(\phi)}}{\sqrt{3}H} \nonumber\\
&&z=\frac{\kappa\phi}{\sqrt{6}} \label{auxilliary},
\end{eqnarray}
and defining
$\lambda\equiv-\frac{1}{\kappa}\frac{V'(\phi)}{V(\phi)}$.

Using these variables, relation (\ref{eq:6a}) gives:
\begin{equation}
\label{Omphi}
 \Omega_{\phi}\equiv\frac{\kappa^{2}\rho_{\phi}}{3H^{2}}=\ve x^2+y^2+\ve 12\xi z
x+\ve 6\xi z^2
\end{equation}
and thus
\begin{equation}
\label{Omm2}
 \Omega_m=1- \Omega_{\phi}=1-(\ve x^2+y^2+\ve 12\xi z
x+\ve 6\xi z^2),
\end{equation}
where $\Omega_{\phi}$ is the scalar field (that is dark energy)
density parameter. Similarly, substituting (\ref{Hdoteff}) into
(\ref{eq:6b}) expressed in terms of $(x,y,z)$ we obtain:
\begin{eqnarray}
\frac{\kappa^{2}p_{\phi}}{3H^{2}}=\ve
(1-4\xi)x^{2}-y^{2}(1+2\sqrt{6}\xi\lambda z) + \ve4\xi x z +
\nonumber\\+ \ve 12 \xi^{2}z^{2} +
\ve6\xi(1-6\xi)z^{2}w_{\text{tot}},\
\end{eqnarray}
and then inserting the above formulae into (\ref{eq:13}) we can
explicitly express $w_{\text{tot}}$, namely {\small{
\begin{equation}
\label{weff}
w_{\text{tot}}=\frac{\ve(1-4\xi)x^{2}-y^{2}(1+2\sqrt{6}\xi\lambda
z) +\ve4\xi x z + \ve12 \xi^{2}z^{2} +
w_{m}\Omega_{m}}{1-\ve6\xi(1-6\xi)z^{2}}.
\end{equation}}}
Inserting this expression for $w_{\text{tot}}$ into
(\ref{Hdoteff}) we can eliminate $\dot{H}$ from
(\ref{eq:6a}),(\ref{eq:6b}) and thus obtain the dark-energy
equation-of-state parameter from (\ref{EoS}) as:
\begin{widetext}
{\small{
\begin{equation}
\label{gammadef} \gamma_{\phi}=1+w_{\phi}=2\, \frac{x^2 [\ve-\ve(2+\ve 3
z^2)\xi+18z^2\xi^2]+4\xi x z [2\ve +9z^2\xi(6\xi-1)]+\xi z
\{-\sqrt{6}y^2\lambda+18\xi z^3(6\xi-1)+\ve 3
z[1+2\xi+y^2(6\xi-1)]\}
 }
 {(x^2+\ve y^2+12\xi x z+6
z^2\xi)[\ve +6 z^2\xi(6\xi-1)]},
\end{equation}}}
\end{widetext}
where for simplicity we have considered the dust case $w_m=0$,
although this is not necessary.

Finally, noting that for any quantity $F$ we have
$\dot{F}=H\frac{dF}{d\ln
 a}$, the cosmological system itself takes the
following autonomous form (see also \cite{Szydlowski:2008in} for
details):
\begin{equation}
\label{sys1} \frac{\ud x}{\ud \ln{a}} = -3x - 12\xi z +\ve
\sqrt{\frac{3}{2}} \lambda y^{2} +  \frac{3}{2} (x+6\xi
z)(1+w_{\text{tot}})
\end{equation}
\begin{eqnarray}
\label{sys2}
 &&\frac{\ud y}{\ud \ln{a}} = -\sqrt{\frac{3}{2}}
\lambda x y + \frac{3}{2} y(1+w_{\text{tot}})\ \ \ \ \ \ \ \ \ \ \
\ \ \ \ \ \ \ \ \ \ \ \ \
\\
\label{sys3a}
&& \frac{\ud z}{\ud \ln{a}} = x\\
\label{sys4a}
 &&\frac{\ud \lambda}{\ud \ln{a}} =
-\sqrt{6}\lambda^{2}\big(\Gamma-1\big)x,
\end{eqnarray}
where $\Gamma=\frac{V'' V}{V'^{2}}$, and $w_{\text{tot}}$ is
determined by (\ref{weff}) with $w_{m}=0$.

The autonomous equations (\ref{sys1})-(\ref{sys4a}) correspond to
the exact cosmological evolution  for non-minimally coupled
quintessence or phantom models, with the dark-energy
equation-of-state parameter $w_{\phi}$ and the dark-energy density
parameter $\Omega_\phi$ given by (\ref{gammadef}) and
(\ref{Omphi}) respectively. We mention that when $\xi=0$, that is
in the minimal-coupling case, all equations coincide with those of
\cite{Scherrer:2007pu,Scherrer:2008be}.

In order to acquire analytic solutions and in particular to obtain
an expression for the experimentally accessible quantity
$w_{\phi}(\Omega_\phi)$, we follow the strategy of
\cite{Scherrer:2007pu,Scherrer:2008be,Setare:2008sf,Sen:2009mc}.
That is, we first differentiate (\ref{gammadef}) and (\ref{Omphi})
with respect to $\ln a$, and making use of
(\ref{sys1})-(\ref{sys4a}) we obtain $\frac{d\gamma_{\phi}}{d\ln
a}$ and $\frac{d\Omega_\phi}{d\ln a}$ as functions of
$x,y,z,\lambda$. Then using (\ref{gammadef}) and (\ref{Omphi}) we
eliminate the auxiliary variables in favor of $\gamma_{\phi}$ and
$\Omega_\phi$, acquiring $\frac{d\gamma_{\phi}}{d\ln a}$ and
$\frac{d\Omega_\phi}{d\ln a}$ as functions of
$\gamma_{\phi},\Omega_\phi$. Finally, we obtain the exact
differential equation for $\frac{d\gamma_{\phi}}{d\Omega_\phi}$ as
\begin{equation}\label{dgamma1}
\frac{d\gamma_{\phi}}{d\Omega_\phi}=\frac{\frac{d\gamma_{\phi}}{d\ln
a}}{\frac{d\Omega_\phi}{d\ln a}}.
\end{equation}
However, comparing to the minimally coupled scenario there is a
difference. In particular, in the former case
\cite{Scherrer:2007pu,Scherrer:2008be,Sen:2009mc} $\gamma_{\phi}$ and
$\Omega_\phi$ depend only on $x$ and $y$, and thus one can easily
acquire the expressions $x(\gamma_{\phi},\Omega_\phi)$ and
$y(\gamma_{\phi},\Omega_\phi)$ in order to eliminate the auxiliary
variables. On the contrary, in the non-minimal coupling case,
where the richer dynamics is reflected in the appearance of an
additional degree of freedom, this procedure is not possible
 since one has to eliminate
three variables ($x$, $y$, $z$) in terms of two ($\gamma_{\phi}$ and
$\Omega_\phi$). Therefore, we make the following assumption: Since
$z$ in all relations appears always multiplied by $\xi$, which is
in general a small quantity, its time variation will not have
significant effects. Moreover, we are considering potentials
to be sufficiently flat satisfying slow-roll conditions (see below).
Under this assumption, the variations of the scalar field  with time
will always be sufficiently small. Thus we replace $z$ by its average value
$z_0$. Doing so, the elimination of $x$ and $y$ in terms of
$\gamma_{\phi}$ and $\Omega_\phi$ can be performed, leading to the
differential equation $\frac{d\gamma_{\phi}}{d\Omega_\phi}=
f(\gamma_{\phi},\Omega_\phi,\lambda)$, which is a rather complicated
expression.

At this stage, similarly to
\cite{Scherrer:2007pu,Scherrer:2008be,Sen:2009mc}, we proceed to
the usual assumptions, based on the nearly flatness of the
potentials, quantitatively expressed as

\begin{eqnarray}
\label{slowroll1} && {1\over{V}}{dV\over{d\phi}}
<< 1\\
\label{slowroll2}
 &&{1\over{V}}{{d^2}V\over{d{\phi^2}}} << 1.
\end{eqnarray}

First, from the definition
$\lambda\equiv-\frac{1}{\kappa}\frac{V'(\phi)}{V(\phi)}$ and using
(\ref{slowroll1}) we conclude that $\lambda$ can approximately
 considered to be a very small constant, so that
\begin{equation}
\lambda\approx\lambda_0=-{1\over{\kappa
V}}{dV\over{d\phi}}\Big|_{\phi=\phi_0}\ll1,
\end{equation}
where $\lambda_0$ and $\phi_0$ are the initial values of $\lambda$
and $\phi$ respectively. This approximation is also consistent
with the $\lambda$-evolution equation (\ref{sys4a}), since
$\frac{\ud \lambda}{\ud \ln{a}}$ is proportional to $\lambda^2$,
i.e it is very small, justifying that $\lambda$ remains almost
constant.
 The second assumption is that, as mentioned in the introduction, $w_{\phi}$ remains close to $-1$
 throughout the evolution, in agreement with observations. Thus,
 $|\gamma_{\phi}|\ll1$, too. Finally, in the non-minimal coupling case at
 hand we make an additional assumption that was not present in
 minimal coupling models
 \cite{Scherrer:2007pu,Scherrer:2008be,Sen:2009mc}. Namely, since
the non-minimal coupling constant $\xi$ is usually a small number,
with the conformal value $\xi=1/6$ being the one that accepts a
reasonable theoretical justification
\cite{nonminimal,Uzan:1999ch,liddle}, we can assume that
$\xi\ll1$, too.

Under these approximations the complicated differential equation
 $\frac{d\gamma_{\phi}}{d\Omega_\phi}=
f(\gamma_{\phi},\Omega_\phi,\lambda)$ acquires the following simple form:
{\small{
\begin{eqnarray}\label{dgamma2}
&&\frac{d\gamma_{\phi}}{d\Omega_\phi}=-\frac{2\gamma_{\phi}}{(1-\Omega_\phi)\Omega_\phi}+
\frac{2}{3}\lambda_0\frac{\sqrt{\ve3\gamma_{\phi}}}{(1-\Omega_\phi)\sqrt{\Omega_\phi}}+\nonumber\\
&&+\xi\left[\frac{8\gamma_{\phi}}{3(1-\Omega_\phi)\Omega_\phi}-
\frac{4}{3}\lambda_0\frac{\sqrt{\ve3\gamma_{\phi}}}{(1-\Omega_\phi)\sqrt{\Omega_\phi}}+\ve
4z_0\,\frac{\sqrt{\ve2\gamma_{\phi}}}{\Omega_\phi^{3/2}}\right].\ \ \ \ \
\end{eqnarray}}}
Note that in the phantom case $\gamma_{\phi}<0$ and $\ve=-1$ while in
quintessence $\gamma_{\phi}>0$, $\ve=1$, thus the combination $\ve\gamma_{\phi}$
is always positive and the aforementioned equation is
well-defined.

Differential equation (\ref{dgamma2}) describes approximately the
cosmological behavior of the model at hand. Note that in the limit
$\xi\rightarrow0$ it coincides with the corresponding one of
\cite{Scherrer:2007pu,Scherrer:2008be}. It can be transformed into
a linear differential equation under the transformation
$s^2=\ve\gamma_{\phi}$ and can be solved exactly. Thus, the general
solution reads:
\begin{widetext}{\small{
\begin{eqnarray}
\label{solusion2}
&&\gamma_{\phi}(\Omega_\phi)=1+w_{\phi}(\Omega_\phi)=\nonumber\\
&&=
\ve\frac{1}{9}\left(\frac{1-\Omega_\phi}{\Omega_\phi}\right)^{2-\frac{8\xi}{3}}
\left\{\ve6\sqrt{2}z_0\xi
\,B\left(\Omega_\phi;\frac{1}{2}-\frac{4\xi}{3},-1+\frac{4\xi}{3}\right)+
[\sqrt{3}\lambda_0(1-2\xi)-\ve6\sqrt{2}z_0\xi]
\,B\left(\Omega_\phi;\frac{3}{2}-\frac{4\xi}{3},-1+\frac{4\xi}{3}\right)
 \right\}^2,\ \ \ \ \ \
\end{eqnarray}}}
\end{widetext}
 where   $B(u;a,b)$ is the incomplete beta function defined as
\begin{equation}
B(u;a,b)=\int_0^u\,t^{a-1}(1-t)^{b-1}dt.
\end{equation}

It can be clearly seen that in the limit $\xi\rightarrow0$
solution (\ref{solusion2}) coincides with the corresponding
expressions of \cite{Scherrer:2007pu,Scherrer:2008be} (for
quintessence and phantom respectively), namely:
\begin{eqnarray}
\label{solusion2xi0} \lim_{\xi\rightarrow0}\gamma_{\phi}(\Omega_\phi)=
\ve\frac{\lambda_0^2}{3}\frac{(1-\Omega_\phi)^2}{\Omega_\phi^2}
\,B\left(\Omega_\phi;\frac{3}{2},-1\right)^2=\nonumber\\
=\ve\frac{\lambda_0^2}{3}\left[\frac{1}{\sqrt{\Omega_\phi}} -
\frac{1}{2}\left(\frac{1}{\Omega_\phi} - 1 \right) \ln
\left(\frac{1+\sqrt{\Omega_\phi}} {1-\sqrt{\Omega_\phi}}
\right)\right]^2.
\end{eqnarray}

In order to express our result in a form more suitable for
comparison with observations, and similarly to
\cite{Scherrer:2007pu,Scherrer:2008be,Setare:2008sf,Sen:2009mc},
we will use the relation
\begin{equation}
\label{Oma} \Omega_\phi(a) = \left[1 + \left(\Omega_{\phi0}^{-1} -
1 \right)a^{-3} \right]^{-1},
\end{equation}
which is the zeroth order solution of $\frac{d\Omega_\phi}{d\ln
a}=g(\Omega_\phi,\gamma_{\phi})$, namely
\begin{equation}
\frac{d\Omega_\phi}{d\ln
a}=3\Omega_\phi(1-\Omega_\phi)+{\cal{O}}(\xi^2)+{\cal{O}}(\gamma_{\phi}^2)+{\cal{O}}(\lambda_0^2)
\label{relations2O},
 \end{equation}
where $\Omega_{\phi0}$ is the value of $\Omega_{\phi}$ at present
($a_0=1$). We mention that (\ref{Oma}) holds only for
$|\gamma_\phi|\ll1$, that is for $w_\phi\approx-1$. Finally, note
that one could equivalently use the redshift $z$ instead of the
scale factor $a$, using $a=(1+z)^{-1}$.

Using (\ref{Oma}) solution (\ref{solusion2}) writes:
\begin{widetext}{\small{
\begin{eqnarray}
\label{solusion3}
 1+w_{\phi}(a)=\ve\frac{1}{9}\left\{
 \frac{\left[1 +
\left(\Omega_{\phi0}^{-1} - 1 \right)a^{-3}
\right](1-\Omega_{\phi0})}{1+(a^3-1)\Omega_{\phi0}}\right\}
^{2-\frac{8\xi}{3}} \left\{\ve6\sqrt{2}z_0\xi \,B\left(\left[1 +
\left(\Omega_{\phi0}^{-1} - 1 \right)a^{-3}
\right]^{-1};\frac{1}{2}-\frac{4\xi}{3},-1+\frac{4\xi}{3}\right)+
\right.\nonumber\\
\left. + [\sqrt{3}\lambda_0(1-2\xi)-\ve6\sqrt{2}z_0\xi]
\,B\left(\left[1 + \left(\Omega_{\phi0}^{-1} - 1 \right)a^{-3}
\right]^{-1};\frac{3}{2}-\frac{4\xi}{3},-1+\frac{4\xi}{3}\right)
 \right\}^2.\ \ \ \
\end{eqnarray}}}
\end{widetext}

Relations (\ref{solusion2}) and  (\ref{solusion3}) are the main
results of the present work. They provide $w_{\phi}(\Omega_\phi)$
and $w_{\phi}(a)$, with $\Omega_{\phi0}$, $\lambda_0$, $z_0$ and
of course $\xi$ as parameters, that is they give the unique,
convergent behavior for the whole sub-class of models where
non-minimally coupled quintessence and phantom fields evolve in
nearly flat potentials. In principle, one can eliminate either
$\lambda_0$ or $z_0$ in terms of $w_{\phi 0}$ (the present value
of $w_{\phi}$). However, contrary to the minimally coupling case
\cite{Scherrer:2007pu,Scherrer:2008be} this straightforward
procedure leads to a complicated expression for $w_{\phi}(a)$, and
thus we prefer to keep (\ref{solusion3}). Finally, we mention that
in the limit $\xi\rightarrow0$ (\ref{solusion3}) coincides with
the corresponding relations of
\cite{Scherrer:2007pu,Scherrer:2008be}.

\section{Cosmological evolution and observational constraints} \label{discuss}

\begin{figure}[ht]
\begin{center}
\mbox{\epsfig{figure=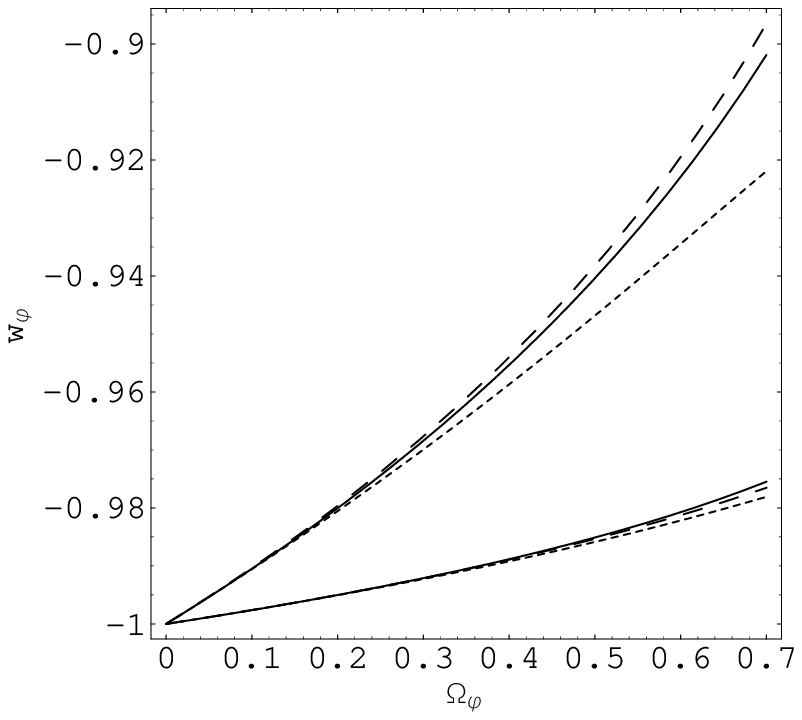,width=8.cm,angle=0}} \caption{{\it
The behavior of $w_{\phi}(\Omega_\phi)$ for the quintessence
scenario in the conformal non-minimal coupling case, i.e with
$\xi=1/6$. The two curve-sets correspond to $\la_0=1$ (upper set)
and $\la_0=0.5$ (lower set). For each set, the dashed and dotted
curves depict the exact numerical evolution for potentials
$V(\phi)=\phi^2$ and $V(\phi)=\phi^{-2}$ respectively, while the
solid curve provides the $w_{\phi}(\Omega_\phi)$-behavior as it
arises from our analytical result (\ref{solusion2}) with $\ve=+1$.
In all cases the present value of $\Omega_{\phi0}$ has been
fitted to be equal to $0.7$ and the initial $w_{\phi}$-value has
been forced to be very close to $-1$ (which yields a relative
small number for $z_0$, of the order of $10^{-5}$).
 }} \label{quintessence1}
\end{center}
\end{figure}

Having obtained approximate analytical solutions, we wish to
explore them, examining their cosmological implications, and to
compare them to the numerically-elaborated exact evolution for a
few different models. In fig.\ref{quintessence1} we present the
behavior of $w_{\phi}(\Omega_\phi)$ for the quintessence scenario,
in the conformal case, i.e with $\xi$ being equal to $1/6$.
The solid curve depicts the $w(\Omega_\phi)$-behavior as it is
given by (\ref{solusion2}) with $\ve=+1$, while the dashed and
dotted curves correspond to the exact evolution for the potentials
$V(\phi)=\phi^2$ and $V(\phi)=\phi^{-2}$ respectively. Finally, we
present the aforementioned plots for two $\la_0$-values, namely
$\la_0=1$ (upper set) and $\la_0=0.5$ (lower set). In all cases
the present value of $\Omega_{\phi0}$ has been fitted to be equal
to $0.7$ and the initial value of $w_{\phi}$ has been forced to be
very close to $-1$, which yields a relative small number for $z_0$
(of the order of $10^{-5}$) and this fact indeed justifies our
assumption to replace $z$ by its average $z_0$ and neglect terms
of ${\cal{O}}(z_0^2)$.

\begin{figure}[ht]
\begin{center}
\mbox{\epsfig{figure=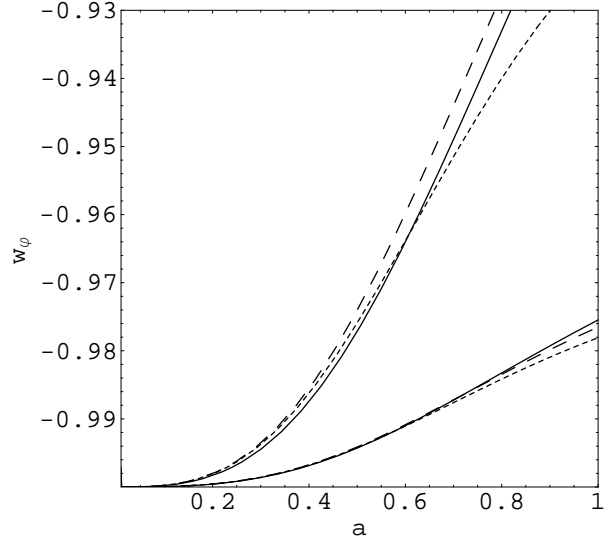,width=8.cm,angle=0}} \caption{ {\it
The behavior of $w_{\phi}(a)$ for the quintessence scenario in the
conformal non-minimal coupling case ($\xi=1/6$). The two
curve-sets correspond to $\la_0=1$ (upper set) and $\la_0=0.5$
(lower set). For each set, the dashed and dotted curves depict the
exact numerical evolution for potentials $V(\phi)=\phi^2$ and
$V(\phi)=\phi^{-2}$ respectively, while the solid curve provides
the $w(a)$-behavior as it arises from our analytical result
(\ref{solusion3}) with $\ve=+1$.  }} \label{quintessence2}
\end{center}
\end{figure}

As expected, agreement is not satisfactory for $\la_0=1$, but it
improves significantly for smaller values. Furthermore, since we
have neglected terms of order $(1+w_{\phi})^2$, the resulting
deviation between the exact evolution and the analytical
expression is larger for larger difference between $w_{\phi}$ and
$-1$. These features are valid for completely different
potentials. We mention that the corresponding deviations are
larger comparing to the minimal coupling case of
\cite{Scherrer:2007pu}, due to the additional errors brought in by
the ${\cal{O}}(\xi^2)$-approximation, but this difference is
relatively weak due to the smallness of the conformal value
($\xi=1/6$). Finally, in fig. \ref{quintessence2} we present the
behavior of $w_{\phi}(a)$ for all the cases of fig.
\ref{quintessence1}, as it arises from (\ref{solusion3}) with
$\ve=+1$. Again we see that our analytical expression is accurate
as long as $w_{\phi}(a)$ is not far from $-1$ and $\la_0$ is
small.

From these two figures we observe that when the potential is
nearly flat, then all the examined models converge to a common
behavior (our analytical solutions (\ref{solusion2}) and
(\ref{solusion3})).
Although non-minimal coupling brings an additional degree of
freedom, i.e an additional parameter, the small value of the
coupling (for instance $\xi=1/6$ in the conformal case) downgrades
significantly the corresponding complexity.

\begin{figure}[ht]
\begin{center}
\mbox{\epsfig{figure=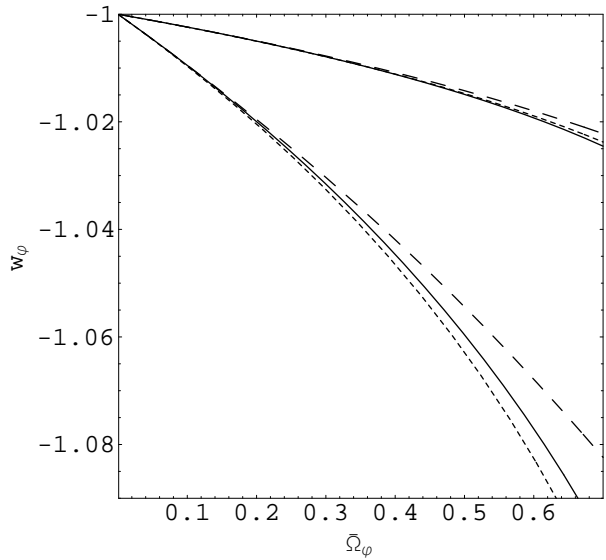,width=8.cm,angle=0}} \caption{{\it
$w_{\phi}(\Omega_\phi)$ for the phantom scenario in the conformal
non-minimal coupling case ($\xi=1/6$). The two curve-sets
correspond to $\la_0=-0.5$ (upper set) and $\la_0=-1$ (lower set).
For each set, the dashed and dotted curves depict the exact
numerical evolution for potentials $V(\phi)=\phi^2$ and
$V(\phi)=\phi^{-2}$ respectively, while the solid curve provides
the $w_{\phi}(\Omega_\phi)$-behavior as it arises from our
analytical result (\ref{solusion2}) with $\ve=-1$. }}
\label{phantom2}
\end{center}
\end{figure}

\begin{figure}[ht]
\begin{center}
\mbox{\epsfig{figure=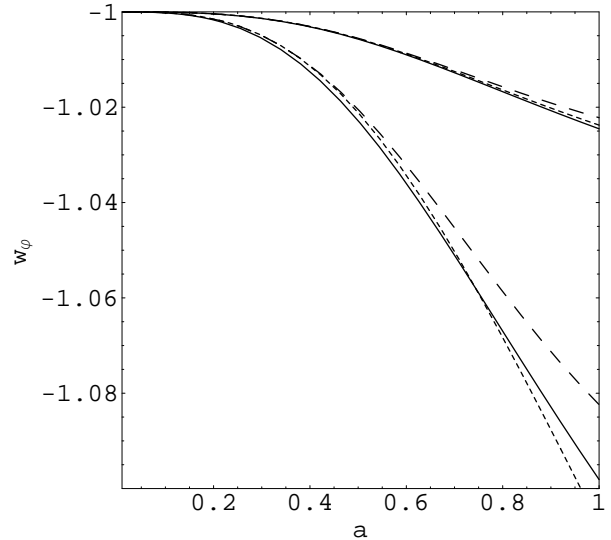,width=8.cm,angle=0}} \caption{ {\it
$w_{\phi}(a)$ for the phantom scenario in the conformal
non-minimal coupling case ($\xi=1/6$). The two curve-sets
correspond to $\la_0=-0.5$ (upper set) and $\la_0=-1$ (lower set).
For each set, the dashed and dotted curves depict the exact
numerical evolution for potentials $V(\phi)=\phi^2$ and
$V(\phi)=\phi^{-2}$ respectively, while the solid curve provides
the $w_{\phi}(a)$-behavior as it arises from our analytical result
(\ref{solusion3}) with $\ve=-1$. }} \label{phantom1}
\end{center}
\end{figure}

In fig. \ref{phantom2} we show  $w_{\phi}(\Omega_\phi)$ for the phantom
scenario with $\xi=1/6$. In this case, and due to the ``inverse''
behavior of phantom fields in potential slopes, we expect $\la$
(defined  as
$\lambda\equiv-\frac{1}{\kappa}\frac{V'(\phi)}{V(\phi)}$) and thus
$\la_0$ too, to be negative. This was also incorporated in the
minimal coupling case \cite{Scherrer:2008be} where the authors
defined $\la$ straightaway with an opposite sign. Furthermore, we
mention that the chosen potentials are not too stiff, in order to
avoid a Big Rip  which is common in phantom scenario
\cite{Kujat,Sami04}.
As we see, the agreement of our analytical expression with the
exact evolution is better for smaller $|\la_0|$ and $|1+w_{\phi}|$. The
same features are observed in the corresponding $w_{\phi}(a)$-behavior
shown in fig. \ref{phantom1}.

\begin{figure}
\begin{center}
\mbox{\epsfig{figure=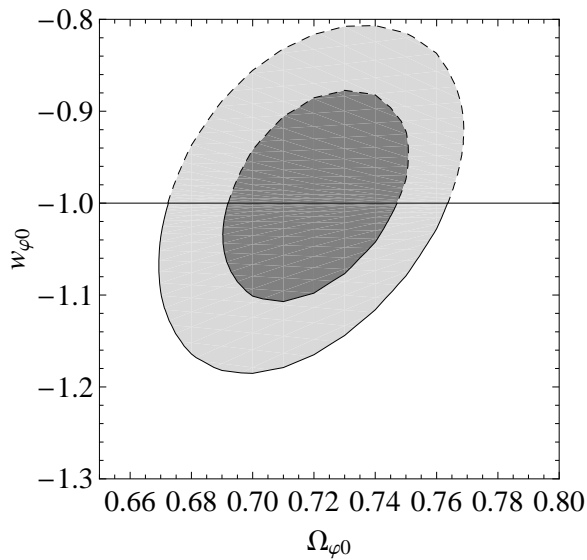,width=7.8cm,angle=0}} \caption{ {\it
$1\sigma$ and $2\sigma$ confidence contours in the plane defined
by the present-day values of $\Omega_{\phi}$ and $w_{\phi}$
(denoted by $\Omega_{\phi0}$ and $w_{\phi0}$ respectively). The
dashed curves correspond to quintessence scenario while the solid
curves to the phantom one. Note that the contours match exactly at
the $w_{\phi0}=-1$ line, although the corresponding
fitting-expressions are not ``symmetric'' between the two cases.
}} \label{contour}
\end{center}
\end{figure}

Having determined the accuracy region of our analytical
expressions, we compare our result for $w_{\phi}(a)$ (given by
(\ref{solusion3})) to supernova Type Ia and Baryon Acoustic
Oscillations data. In this expression for $w_{\phi}(a)$ there are
four parameters, namely $\Omega_{\phi0}$, $\xi$, $z_{0}$ and
$\lambda_{0}$. As we have mentioned, we have assumed the conformal
coupled case ($\xi = 1/6$) and we have considered $z_{0}$ to be of
the order of $10^{-5}$ for our present investigation. While taking
different values for $\xi$ will definitely change our result, our
conclusions below are not so sensitive on $z_{0}$ as long as it is
sufficiently small. This is also one crucial requirement for our
analytical result to be valid. With two remaining parameters
$\Omega_{\phi0}$ and $\lambda_{0}$, one can relate $\lambda_{0}$
to the present day value of the equation of state $w_{\phi0}$.
Hence our aim is to put constraints on the model parameters
$w_{\phi0}$ and $\Omega_{\phi0}$.
 At present, we have the Union08 compilation of
the SnIa data which contains around 307 data points
\cite{kowalski}. This is world's published first heterogeneous SN
data set containing large sample of data from SNLS, Essence
survey, high redshift supernova data from Hubble Space telescope,
as well as several small data sets. We use this data set together
with the BAO data from SDSS (Sloan Digital Sky Survey) \cite{bao}.
The corresponding confidence contours are depicted in fig.
\ref{contour}.

We should mention that while fitting our model with the
observational data, we have to do it separately for quintessence
and phantom case, since the expressions for $(1+w_{\phi})$ and
$-(1+w_{\phi})$ as a function of scale factor are not the same.
This is unlike the minimally coupled case
\cite{Scherrer:2007pu,Scherrer:2008be}, where they are exactly
identical. In this regard, and from the fitting procedure point of
view, one does not expect in general to have the confidence
contours for quintessence and phantom case to match exactly to
give a complete ellipse, simply because we are fitting two
different $w_{\phi}(a)$. On the other hand, since we have switched
the sign of the kinetic term of the scalar field in order to
obtain the phantom model, we theoretically expect continuity for
the exact solution and the contour to cross smoothly to the
phantom region. Thus, the fact that the contours match exactly, as
shown in fig. \ref{contour}, indicates that our approximate method
works well for both quintessence and phantom cases.

As we observe, and similarly to the minimally-coupled cases
\cite{Scherrer:2007pu,Scherrer:2008be,amna}, in order to acquire
tight constraints on $w_{\phi0}$ one has to impose strong bounds
on $\Omega_{\phi0}$, arising from the corresponding constraints on
$\Omega_{m0}$ from structure formation data like the shape
parameter for the matter power spectrum. However, with the present
data, and although the behavior is richer than the minimally
coupling scenarios, one cannot distinguish between the present
model and a $\Lambda$CDM one.

\section{Conclusions}
 \label{concl}

In the present work we have studied dark energy scenarios in which
the scalar field is non-minimally coupled to gravity and it
evolves in nearly flat potentials in order for the slow-roll
conditions to be satisfied. For completeness, we have investigated
both quintessence and phantom cases in a unified way, having in
mind that the later paradigm could have problems at the quantum
level (the relevant discussion is still open in the literature
since for instance in \cite{Cline:2003gs} the authors reveal the
causality and stability problems while in \cite{quantumphantom0}
the authors construct a phantom theory consistent with the basic
requirements of quantum field theory with the phantom fields
arising as an effective description).

We have shown that all such models converge to a common behavior
and we have extracted the corresponding approximate analytical
expressions for $w_{\phi}(\Omega_\phi)$ (relation
(\ref{solusion2})) and for $w_{\phi}(a)$ (relation
(\ref{solusion3})). The smallness of the non-minimal coupling
parameter ($\xi=1/6$ in the conformal case) enhances the validity
of our formulae, since it downgrades the role of the additional
degree of freedom comparing to the minimal-coupling case. In
particular, the errors are typically within $2\%$, and are smaller
for $w_{\phi}$ being closer to $-1$ and for more flat potentials.
Moreover, a comparison with observations using Baryon Acoustic
Oscillation and latest Supernovae data reveals that the situation
is similar to that of a minimally coupled scalar field. That is,
under the present data the model cannot be distinguished from a
$\Lambda$CDM scenario. However, one should note that in our
investigation we have fixed the parameter $\xi $ to its conformal
value $1/6$. Keeping $\xi$ arbitrary will make the parameter space
broader, which in turn will lead further to less stringent
constraints. In other words, switching on non-minimal coupling it
becomes easier to fit the observational data, at least at the
background cosmological level.

Non-minimal coupling leads to richer cosmological behavior
comparing to the minimal scenario. For instance, although in
minimal coupling with completely flat potentials one obtains the
cosmological constant universe (i.e $w=-1=const.$)
\cite{Scherrer:2007pu,Scherrer:2008be}, in the model at hand, even
in this case one obtains a varying $w$ resulting purely from
non-minimal coupling. In addition, it is interesting to see that
non-minimal coupling introduces an ``asymmetry'' between
quintessence and phantom cases, which is visible straightaway in
the energy density and pressure definitions (\ref{eq:6a}) and
(\ref{eq:6b}). However, even if the expressions are non-symmetric
in the two sides of the phantom divide, the observation comparison
reveals a non-trivial match of the confidence contours at the
$w_{\phi0} = -1$ line. This feature provides a physical self-consistency test
of the obtained analytical expressions of the present work.

Finally, we mention that we have only considered observational
data related to distance measurements, depending only on the
background homogeneous and isotropic universe. An additional
consideration of the growth of matter perturbations in these
models, can have interesting effects and may help to distinguish
between the present model and the cosmological-constant one. This
will be our aim in a future investigation.
\\

\paragraph*{{\bf{Acknowledgements:}}}
ENS wishes to thank Institut de Physique Th\'eorique, CEA, for the
hospitality during the preparation of the present work. AAS
acknowledges the financial support provided by the University
Grants Commission, Govt. Of India through the major research
project grant (Grant No:33-28/2007(SR)). AAS also knowledges the
financial grant provided by the Theory Divison at CERN and the
High Energy Physics Division at Abdus Salam International Center
For Theoretical Physics, where part of the work has been done.


\begin{thebibliography}{99}



\bibitem{obs1}A. G. Riess et al., Astron. J. {\bf 116}, 1009 (1998);
S. Perlmutter et al., Astrophys. J. {\bf 517}, 565 (1999); J. L.
Tonry et al., Astrophys. J. {\bf 594}, 1 (2003).


\bibitem{cmbr}A. Melchiorri et al., Astrophys. J. Lett. {\bf 536},
L63 (2000); A. E. Lange et al., Phys. Rev. D {\bf 63}, 042001
(2001); A. H. Jaffe et al., Phys. Rev. Lett. {\bf 86}, 3475
(2001); C. B. Netterfield et al., Astrophys. J. {\bf 571}, 604
(2002); N. W. Halverson et al., Astrophys. J. {\bf 568}, 38
(2002).

\bibitem{wmap}S. Bridle, O. Lahab, J. P. Ostriker and P. J. Steinhardt,
                 Science {\bf 299}, 1532 (2003); C. Bennett et al.,
Astrophys. J. Suppl. Ser. {\bf 148}, 1 (2003); G. Hinshaw et al.,
Astrophys. J. Suppl. Ser. {\bf 148}, 135 (2003); A. Kogut et al.,
Astrophys. J. Suppl. Ser. {\bf 148}, 161 (2003);
              D. N. Spergel et al., Astrophys. J. Suppl. Ser. {\bf 148}, 175 (2003).


\bibitem{ordishov}
 P. Bin\'{e}truy, C. Deffayet, D. Langlois, Nucl. Phys. B {\bf 565}, 269 (2000);
G.R. Dvali, G. Gabadadze, M. Porrati, Phys. Lett. B {\bf 485}, 208
(2000); S. Capozziello, Int. J. Mod. Phys. D {\bf 11}, 483 (2002);
 S.Nojiri
and S.~D.~Odintsov, Phys. Rev. D {\bf{68}}, 123512 (2003);
  P.~S.~Apostolopoulos, N.~Brouzakis, E.~N.~Saridakis and N.~Tetradis,
  Phys.\ Rev.\  D {\bf 72}, 044013 (2005);
S.Nojiri and S.~D.~Odintsov, Int. J. Geom. Meth. Mod. Phys.
{\bf{4}}, 115 (2007);
  F.~K.~Diakonos and E.~N.~Saridakis,
 JCAP {\bf 0902}, 030 (2009);
  E.~N.~Saridakis,
  arXiv:0905.3532 [hep-th].


\bibitem{quintess}
 P.~J.~E.~Peebles and  B.~Ratra, \apj {\bf 325}, L17 (1988);
C.~Wetterich, Nucl.\ Phys.\ B {\bf 302}, 668 (1988); M. S. Turner
and M. White, Phys. Rev. D {\bf{56}}, 4439 (1997); R. R. Caldwell,
R. Dave and P. J. Steinhardt, Phys. Rev. Lett. {\bf{80}}, 1582
(1998);  I.~Zlatev, L.~M.~Wang and P.~J.~Steinhardt, Phys.\ Rev.\
Lett.\ {\bf 82}, 896 (1999);
  A.~J.~Albrecht, C.~P.~Burgess, F.~Ravndal and C.~Skordis,
  Phys.\ Rev.\  D {\bf 65}, 123507 (2002);   M.~Sahlen, A.~R.~Liddle and D.~Parkinson,
  Phys.\ Rev.\  D {\bf 72}, 083511 (2005);   M.~Sahlen, A.~R.~Liddle and D.~Parkinson,
  Phys.\ Rev.\  D {\bf 75}, 023502 (2007);   D.~Huterer and H.~V.~Peiris,
  Phys.\ Rev.\  D {\bf 75}, 083503 (2007);
 Z.~K.~Guo, N.~Ohta and Y.~Z.~Zhang,
Mod.\ Phys.\ Lett.\  A {\bf 22}, 883 (2007);
  S.~Dutta, E.~N.~Saridakis and R.~J.~Scherrer,
Phys.\ Rev.\  D {\bf 79}, 103005 (2009) [arXiv:0903.3412
[astro-ph.CO]].

\bibitem{phant} R. R. Caldwell, Phys.
Lett. B {\bf{545}}, 23 (2002); R.~R.~Caldwell, M.~Kamionkowski and
N.~N.~Weinberg, Phys. Rev. Lett. {\bf 91}, 071301 (2003);
  P.~F.~Gonzalez-Diaz and C.~L.~Siguenza,
  Nucl.\ Phys.\  B {\bf 697}, 363 (2004);
S. Nojiri and S. D. Odintsov, Phys. Rev. D  {\bf{72}}, 023003
(2005); H. Garcia-Compean, G. Garcia-Jimenez,  O. Obregon, and C.
Ramirez, JCAP {\bf 0807}, 016 (2008);
  X.~m.~Chen, Y.~g.~Gong and E.~N.~Saridakis,
  JCAP {\bf 0904}, 001 (2009);
  E.~N.~Saridakis,
  Nucl.\ Phys.\  B {\bf 819}, 116 (2009).



\bibitem{quintom}
B.~Feng, X.~L.~Wang and X.~M.~Zhang, Phys.\ Lett.\  B {\bf 607},
35 (2005);
Z. K. Guo, {\it{et al.}}, Phys. Lett. B {\bf 608}, 177 (2005);
M.-Z Li, B. Feng, X.-M Zhang, JCAP,  {\bf 0512}, 002 (2005);
  Y.~f.~Cai, H.~Li, Y.~S.~Piao and X.~m.~Zhang,
  Phys.\ Lett.\  B {\bf 646}, 141 (2007);
 W.
Zhao and Y. Zhang, Phys. Rev. D {\bf73}, 123509 (2006);
  M.~R.~Setare and E.~N.~Saridakis,
  Phys.\ Lett.\  B {\bf 668}, 177 (2008);
  M.~R.~Setare and E.~N.~Saridakis,
  JCAP {\bf 0809}, 026 (2008).


\bibitem{quinteflat}
K. Griest, \prd {\bf 66}, 123501 (2002); S. Bludman, \prd {\bf
69}, 122002 (2004); R.R. Caldwell and E.V. Linder, \prl {\bf 95},
141301 (2005); E.V. Linder, \prd {\bf 73}, 063010 (2006); S.
Chongchitnan and G. Efstathiou, \prd {\bf 76}, 043508 (2007).



\bibitem{Kujat}
  J.~Kujat, R.~J.~Scherrer and A.~A.~Sen,
  Phys.\ Rev.\  D {\bf 74}, 083501 (2006).




\bibitem{Huang}
  Q.~G.~Huang,
  Phys.\ Rev.\  D {\bf 77}, 103518 (2008);
  E.~N.~Saridakis,
Phys.\ Lett.\  B {\bf 676}, 7 (2009).



\bibitem{Scherrer:2007pu}
  R.~J.~Scherrer and A.~A.~Sen,
  Phys.\ Rev.\  D {\bf 77}, 083515 (2008).


\bibitem{Scherrer:2008be}
  R.~J.~Scherrer and A.~A.~Sen,
  Phys.\ Rev.\  D {\bf 78}, 067303 (2008).


\bibitem{Setare:2008sf}
  M.~R.~Setare and E.~N.~Saridakis,
  Phys.\ Rev.\  D {\bf 79}, 043005 (2009).

\bibitem{amna}
  A.~Ali, M.~Sami and A.~A.~Sen,
  arXiv:0904.1070 [astro-ph.CO].

\bibitem{Uzan:1999ch}
  V.~Sahni and S.~Habib,
  Phys.\ Rev.\ Lett.\  {\bf 81}, 1766 (1998);
J.~P.~Uzan, Phys.\ Rev.\ D {\bf 59}, 123510 (1999);
 A. A. Sen
and S. Sen, Mod. Phys. Lett. A {\bf 16}, 1303 (2001);
 S. Sen and
A. A. Sen, Phys. Rev. D {\bf 63}, 124006 (2001);
 T. Chiba, Phys.
Rev. D {\bf 60}, 083508 (1999);
 F.~Perrotta, C.~Baccigalupi and S.~Matarrese,
  Phys.\ Rev.\  D {\bf 61}, 023507 (2000);
E. Elizalde, S. Nojiri and S. Odintsov, Phys. Rev. D {\bf 70},
043539 (2004); V. K. Onemli and R. P. Woodard, Phys. Rev. D {\bf
70}, 107301 (2004);  E. Elizalde {\it et al}, Phys. Rev. D {\bf
77}, 106005 (2008).


\bibitem{nonminimal}
V.~Faraoni, Phys. Rev. D {\bf{68}}, 063508 (2003);
 S. Nojiri, S.~D.~Odintsov
and M.~Sami, Phys. Rev. D {\bf{74}}, 046004 (2006);  M.
Szydlowski, O. Hrycyna and A. Kurek,
 Phys. Rev. D {\bf 77}, 027302 (2008);  O. Hrycyna, M. Szydlowski, Phys. Rev. D {\bf 76},
 123510 (2007);
  M.~R.~Setare and E.~N.~Saridakis,
  Phys.\ Lett.\  B {\bf 671}, 331 (2009);
  M.~R.~Setare and E.~N.~Saridakis,
    JCAP {\bf 0903}, 002 (2009).

\bibitem{liddle}
A. R. Liddle and R. J. Scherrer, Phys. Rev. D {\bf 59},
023509 (1998);
  V.~Faraoni,
  Phys.\ Rev.\  D {\bf 62}, 023504 (2000);
  N.~Bartolo and M.~Pietroni,
  Phys.\ Rev.\  D {\bf 61}, 023518 (2000);
O. Bertolami and P. J. Martins, Phys. Rev. D {\bf 61}, 064007
(2000); R. de Ritis, A. A. Marino, C. Rubano and P. Scudellaro,
Phys. Rev. D. {\bf 62}, 043506 (2000);
 T.D. Saini, S. Raychaudhury, V. Sahni and A. A. Starobinsky, Phys.
Rev. Lett. {\bf 85}, 1162 (2000);
 S.~Sen and T.~R.~Seshadri,
  Int.\ J.\ Mod.\ Phys.\  D {\bf 12}, 445 (2003).


\bibitem{Sen:2009mc}
  A.~A.~Sen, G.~Gupta and S.~Das,
  arXiv:0901.0173 [astro-ph.CO].


\bibitem{Copeland:1997et}
P.G. Ferreira, M. Joyce, Phys. Rev. Lett.  {\bf79}, 4740 (1997);
  E.~J.~Copeland, A.~R.~Liddle and D.~Wands,
  Phys.\ Rev.\  D {\bf 57}, 4686 (1998).


\bibitem{Szydlowski:2008in}
  M.~Szydlowski and O.~Hrycyna,
  JCAP {\bf 0901}, 039 (2009).





\bibitem{Sami04} M. Sami, A. Toporensky, Mod. Phys. Lett. A {\bf 19}, 1509 (2004).



\bibitem{kowalski}
M.~Kowalski et.al, Astrophys.\ J. {\bf 686}, 749 (2008).


\bibitem{bao}
D.~J.~Eisenstein et.al, Astrophys.\ J. {\bf 633}, 560 (2005).



    \bibitem{Cline:2003gs}
    J.~M.~Cline, S.~Jeon and G.~D.~Moore,
    Phys.\ Rev.\  D {\bf 70}, 043543 (2004).

\bibitem{quantumphantom0}
  S.~Nojiri and S.~D.~Odintsov,
  Phys.\ Lett.\  B {\bf 562}, 147 (2003);
  S.~Nojiri and S.~D.~Odintsov,
  Phys.\ Lett.\  B {\bf 571}, 1 (2003).


\end{thebibliography}
\end{document}